\documentclass[aps,reprint,floatfix]{revtex4-1}
\usepackage{amsmath}    
\usepackage{graphicx}  
\usepackage{verbatim}  
\usepackage{color}      
\usepackage{subfigure} 
\usepackage{hyperref}   
\usepackage{xcolor}
\begin{document}

\title{Thermalization of hot electrons via interfacial electron-magnon interaction }
\author{Subrata Chakraborty$^{1,2}$}
\email[Correspondence to: ]{schrkmv@gmail.com}
 \author{Tero T. Heikkil{\"a}$^1$}
\email{tero.t.heikkila@jyu.fi}
\affiliation{$^1$Department of Physics and Nanoscience Center, University of Jyv{\"a}skyl{\"a},
 P.O. Box 35 (YFL), FI-40014 Jyv{\"a}skyl{\"a}, Finland}
  \affiliation{$^2$Department of Physics, Queens College of the City University of New York, Queens, NY 11367, USA}

\date{\today}

\begin{abstract} 
Recent work on layered structures of superconductors (S) or normal
metals (N) in contact with ferromagnetic insulators (FI) has shown how
the properties of the previous can be strongly affected by the
magnetic proximity effect due to the static FI magnetization. Here we
show that such structures can also exhibit a new electron
thermalization mechanism due to the coupling of electrons with the dynamic magnetization, i.e., magnons in FI. We here study the heat flow between the two systems and find that in thin films the heat conductance due to the interfacial electron-magnon collisions can dominate over the well-known electron-phonon coupling below a certain characteristic temperature that can be straightforwardly reached with present-day experiments. We also study the role of the magnon band gap and the induced spin-splitting field induced in S on the resulting heat conductance and show that heat balance experiments can reveal information about such quantities in a way quite different from typical magnon spectroscopy experiments.
\end{abstract}

\maketitle

\section{Introduction}

\begin{figure}[h]
\includegraphics[width=8cm]{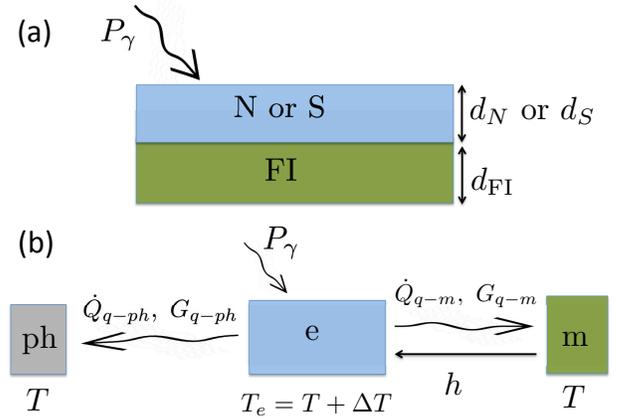}
\caption{\label{fig1} (a) Schematic of a hybrid bilayer of a normal metal (N) or a superconductor (S)
of thickness $d_N$ or $d_S$ in a good contact with a ferromagnetic insulator (FI) of thickness $d_{\mathrm{FI}}$. 
(b) $P_\gamma$ denotes
the incident radiation power, which increases the electronic temperature by an amount $\Delta T$ from some initial temperature $T$ dictated by the temperature of the baths. For a low and constant power $P_\gamma$, the magnitude of $\Delta T = P_\gamma/G_{\rm th}^{\rm tot}$ is dictated by the total heat conductance, $G_{\rm th}^{\rm tot}$, to the heat bath. This heat conductance is typically due to the coupling of the electrons (e) to the phonons (ph), 
but at a low enough temperature also magnons (m) in the FI film start to be relevant, and eventually may become the dominant heat conduction mechanism. 
Here $h$ denotes the spin-splitting field possibly induced to N or S via the magnetic proximity effect. $\dot{Q}_{\rm q-ph}$ and $\dot{Q}_{\rm q-m}$ 
stand for the rates of heat flow due to electron-phonon interaction and 
electron-magnon collisions, respectively and $G_{\rm q-ph}$ and $G_{\rm q-m}$ are the corresponding heat conductances.}
\end{figure}

The progress in low temperature solid state device technology, such as thermometry and electromagnetic radiation detection 
\cite{TEScitation,Grossman, Bluzer, Sergeev, Giazotto, Govenius,Tero_2006,Tero_bolometer,Tero_calorimeter},
electron refrigeration \cite{Kawabata2013,Rouco2018}
and new solutions for quantum information processing \cite{Lian},
call for an improved understanding of the thermalization mechanisms. This is particularly relevant at their usual sub-Kelvin operating temperatures and in hybrid structures. We schematically represent an example hybrid structure in Fig.~\ref{fig1}, based on a thin-film normal metal (N) or a thin-film 
superconductor (S) in contact with a thin-film ferromagnetic insulator (FI). It can be a part of some low-temperature 
thermometric device, such as a thermoelectric radiation detector (TED) \cite{Tero_bolometer,Tero_calorimeter}. When such devices are operated, they are often brought out of equilibrium via a process involving absorption of an electromagnetic field with power $P_\gamma$. This power may be the one under study as in radiation detectors, or one inadvertently brought in when operating the device. As schematized in Fig.~\ref{fig1}, this power initially heats up the electrons of the N or S,  and then the hot electrons dissipate the heat via coupling to larger heat baths, typically via coupling to the phonons (ph)
\cite{wellstood,Tero_2006,Kopnin,Maisi,Tero_2017}. In systems with ferromagnetic elements, such as the one shown in Fig.~\ref{fig1}, the electrons can also couple to the magnons, which can then conduct the energy away from the heated region. This mechanism we study in this paper.

The interfacial electron-magnon interaction strength can be quite large, 
and hence important for the thin film materials, as the recent work on superconductivity induced 
in a metal due to interfacial electron-magnon interaction \cite{Rohling_2018},
and spin transport across normal metal and ferromagnetic insulator \cite{Bender_2012, Takahashi_2010} suggest.
There have been various research works, such as spin pumping, spin and charge tunneling current
in magnetic multilayered structures \cite{Bauer,Branislav,Cheng_2017,Yaroslav,Schreier}, 
which one can also independently analyze via interfacial electron-magnon interaction. 
In this work we demonstrate that the electron-magnon heat flow can be as important as electron-phonon heat 
conduction below a certain characteristic temperature, for a certain regime of electron-magnon interaction strength, magnon band gap and 
spin-splitting field. At high temperatures electron-phonon heat flow dominates over electron-magnon heat flow. 
The dominance of the interfacial electron-magnon heat flow over electron-phonon heat transport in the bulk below a
characteristic temperature is due to the difference of the magnon and phonon dispersions, and hence the
dissimilarity in the magnon density of states at the N-FI or S-FI interface and the phonon density of states in the bulk.

Our present work is  especially important in the context of proposals for a new kind of a low temperature thermoelectric 
radiation detector (TED) \cite{Tero_bolometer,Tero_calorimeter}, which can rival the contemporary device technologies, 
such as transition edge sensor (TES) and kinetic inductance detector (KID) \cite{TEScitation,Grossman, Bluzer, Sergeev, Giazotto, Govenius}. 
The TED is based on a combination of a thin film spin-split superconductor with a spin-polarized tunnel junction, and it utilizes the recently discovered giant
thermoelectric effect in superconductor-ferromagnet hybrid structures \cite{Ozaeta_prl,machon2013nonlocal,machon2014giant,Kolenda,Kolendaprb2017,rezaei2018spin,Tero_2017}. Spin-split superconductors can also be used to generate different types of devices combining thermoelectricity and the macroscopic phase coherence of the superconducting state \cite{giazotto2014proposal,giazotto2015}. 
One way to realize the spin-split superconductor is to couple the S with FI.
Such devices are the most sensitive at the lowest temperatures reached. This is why understanding the 
thermalization mechanisms directly improves the design of such devices.
 
In what follows we first present the theory of electron-magnon heat transport in N-FI and S-FI contacts. Then we discuss our
results on electron-magnon heat conductance and compare it with electron-phonon heat conductance of N or S
to establish the regime where the previous dominates. 

\section{Theoretical model}

To study the heat conduction due to interfacial electron-magnon collisions in N-FI or S-FI hybrid structures, 
we consider the effective model Hamiltonian 
\begin{eqnarray}
\hat{H} &=& \hat{H}_e +\hat{H}_{em} + \hat{H}_m \label{em1}, \\
\hat{H}_e &=& \sum_{\vec{k}\sigma} (\epsilon_{\vec{k}\sigma}-\mu) c^\dag_{\vec{k}\sigma}c_{\vec{k}\sigma}~~~~~~~\mathrm{for~N}, \nonumber  \\
&=& \sum_{\vec{k}\sigma} E_{\vec{k}\sigma} \gamma^\dag_{\vec{k}\sigma}\gamma_{\vec{k}\sigma}~~~~~~~~~~~~~\mathrm{for~S} \label{em2}, \\
\hat{H}_{m} &=& \sum_{\vec{q}} \omega_{\vec{q}} a^\dag_{\vec{q}}a_{\vec{q}}  \label{em3}, \\
\hat{H}_{em}  &=& -g_{em} \sum_{\vec{k},~\vec{q}} 
c^\dag_{\vec{k}\uparrow}c_{\vec{k}+\vec{q}\downarrow}a^\dag_{\vec{q}} + \mathrm{h.c}.    \label{em4}
\end{eqnarray}
Here $\hat{H}_e$, $\hat{H}_m$ and $\hat{H}_{em}$ 
 stand for the Hamiltonian of the quasi two-dimensional (thin film) N or S, for the quasi two-dimensional FI  \cite{vanvleck}
and the electron-magnon interaction due to the N-FI or S-FI contact \cite{Rohling_2018}. We assume low enough temperatures so that the thickness $d_{N/S}$
satisfies $d_{N/S} \ll 2\pi\hbar v_F/(k_BT)$,
where $v_F$ is the Fermi velocity of the electronic system and $T$ is the temperature. In this case the thin films can be considered effectively two-dimensional and the sums over $\vec{k},\vec{q}$ in Eqs.~\eqref{em2}-\eqref{em4} are also two-dimensional.

In Eq.~\eqref{em2} we denote the electron energy  $\epsilon_{\vec{k}\sigma}=\epsilon_{\vec{k}}-h\sigma$ for spin $\sigma$, 
where $\sigma =\pm 1$ for $\sigma =\uparrow/\downarrow$,
and the spin dependent Bogoliubon energy $E_{\vec{k}\sigma}=E_{\vec{k}}-h\sigma$. 
Hence $\epsilon_{\vec{k}}=\mu+\hbar v_F(k-k_F)$ and $E_{\vec{k}}=\sqrt{(\epsilon_{\vec{k}}-\mu)^2+\Delta^2}$,
where $\mu$ and $k_F$ are the chemical potential and the magnitude of the Fermi wave vector in the N or S.  $\Delta$ is the superconducting gap of S and $h$ stands for the spin-splitting field exerted on N or S due to FI.
In Eq.~\eqref{em3}  $\omega_{\vec{q}}=\omega_0+Bq^2$ represents the magnon energy dispersion relation in FI, 
with $\omega_0 \geq 0$ and $B=J_{\mathrm{ex}}zs\zeta^2/2>0$ \cite{vanvleck}, where $J_{\mathrm{ex}}$, $s$, $z$ and $\zeta$
are the isotropic exchange coupling energy, effective lattice spin, coordination number and the lattice constant of FI, respectively. 
The effective electron-magnon coupling energy  is defined as $g_{em}\sqrt{A}=-J\zeta\sqrt{2s}$, 
where $A$ is the area of the contact surface,  $J$ is the exchange energy between the
electrons of the N or S with FI and $\zeta^2=A/N_0$ where $N_0$ is the number of lattice points of FI at the surface of the N-FI or S-FI contact.
In Eqs.~\eqref{em2}-\eqref{em4} $c$, $\gamma$ and $a$ are the annihilation operators for the electrons of the N or S, 
Bogoliubon operator of S and magnon operator for FI, respectively. For S we have 
$c_{\vec{k}\sigma}=v_{\vec{k}\sigma}\gamma^\dag_{-\vec{k}-\sigma}+u^*_{\vec{k}\sigma}\gamma_{\vec{k}\sigma}$, 
$\left|u_{\vec{k}\sigma}\right|^2+\left|v_{\vec{k}\sigma}\right|^2=1$, 
$u_{\vec{k}\uparrow}=u_{\vec{k}}$, $u_{\vec{k}\downarrow}=u_{\vec{k}}$,
$v_{\vec{k}\uparrow}=v_{\vec{k}}$ and $v_{\vec{k}\downarrow}=-v_{-\vec{k}}$, 
$\Delta^*v_{\vec{k}}/u_{\vec{k}}=E_{\vec{k}}-(\epsilon_{\vec{k}}-\mu)$,
$2\left|v_{\vec{k}}\right|^2=\left[ 1-\left({\epsilon_{\vec{k}}-\mu}\right)/{E_{\vec{k}}}\right]$.

Using the Hamiltonians of Eqs.~\eqref{em1}-\eqref{em4}, we calculate the heat flow from the magnons of the FI to the electrons 
of the N or S, according to the Kubo linear response theory, as
$\left<\dot{\hat{H}}_m(t)\right> = \left<\dot{\hat{H}}_m(t)\right>_0 -\frac{i}{\hbar} \int_{-\infty}^t dt^\prime 
\left< \left[\dot{\hat{H}}_m(t),\hat{H}_{em} (t^\prime) \right]\right>_0$ \cite{Kubo}. Here $\left<\cdot\cdot\cdot \right>_0$ 
stands for thermal averaging over the non-interacting system. As a result we obtain the rate of heat flow
from the magnons to the electrons as
\begin{align}
\lim_{t\to\infty} \left<\dot{\hat{H}}_m(t)\right> &= \Lambda \left(Q_{\rm q-m}^{(+)}- Q_{\rm q-m}^{(-)}\right),
\label{emag1} \\
 \Lambda &=\frac{(k_F^2J\zeta\sqrt{2s})^2}{(16\pi^2\hbar\mu^2)} \label{march73}
\end{align} 
and
\begin{eqnarray}
 Q_{\rm q-m}^{(\pm)} & =& 2 \int_{-\infty}^{\infty} dE \int_{\omega_0}^{4\omega_F+\omega_0} d\omega 
\left[n(\omega,T_e) - n(\omega,T)\right]   \nonumber \\
&& \times K^{(\pm)}(E,\omega) 
\left[ f(E\mp\omega,T_e) - f(E,T_e) \right],  
\label{emag2}
\end{eqnarray}
where $n(x,T)=\left[\exp[x/(k_BT)] -1\right]^{-1}$ and $f(x,T)=\left[\exp[x/(k_BT)] +1\right]^{-1}$
are the Bose-Einstein and Fermi-Dirac distributions, respectively.  $\omega_F=Bk_F^2$ is the magnon equivalent of the
Bloch-Gr\"uneisen energy, originating from the requirement for simultaneous energy and momentum conservation. $T$ and $T_e$
are the temperatures of the magnons and electrons. The matrix element of the coupling results into the kernel terms $K^{(\pm)}(E,\omega)$, which are given below for normal and superconducting metals coupled to the ferromagnetic insulator [Eqs.~\eqref{emag5} and \eqref{em6c}, respectively].
Finally let us obtain the electron-magnon heat conductance, $G_{\rm q-m}$, within linear response $\Delta T = T_e-T \ll T$ as
\begin{eqnarray}
G_{\rm q-m} = \lim_{t\to\infty} \frac{\left<\dot{\hat{H}}_m(t)\right>}{\Delta T} 
= \Lambda A \left(G_{\rm q-m}^{(+)}- G_{\rm q-m}^{(-)}\right), 
\label{emag3}
\end{eqnarray} 
where 
\begin{eqnarray}
  G^{(\pm)}_{\rm q-m} &=&  \frac{1}{2 k_BT^2} \int_{-\infty}^{\infty} dE \int_{\omega_0}^{4\omega_F+\omega_0} d\omega~  \omega
   \sinh^{-2}\left(\frac{\omega}{2k_BT}\right)  \nonumber \\
& \times & K^{(\pm)}(E,\omega)   \left[ f(E\mp \omega,T)-f(E,T) \right].
\label{emag4} 
\end{eqnarray}
The steps leading to Eqs.~\eqref{emag1}-\eqref{emag4} correspond to the Born approximation similar to the one used for studying electron-phonon heat transport in earlier works \cite{wellstood,Kopnin,Maisi}.

In what follows we assume $k_BT$, $h$, $\omega_{\vec{q}}$ and $\omega_F \ll \mu$, where $\omega_{\vec{q}}$ are the relevant 
magnons at low temperatures.
As a result, we obtain the kernel term for N-FI as (see the  discussion in Appendix \ref{DCT}) $K^{(\pm)}(E,\omega) = K(\omega)$ with
\begin{eqnarray}
&&  K (\omega) = \sqrt{\frac{\omega}{\omega_F}}\sqrt{\frac{\omega}{\omega -\omega_0}}    
\left[4-\left(\frac{\omega - \omega_0}{\omega_F}\right)\right]^{-1/2}. ~~~~
 \label{emag5}
\end{eqnarray}
Since the kernel is independent of $E$, we can perform the integral over $E$ in Eq.~\eqref{emag3} and obtain
\begin{eqnarray}
    G_{\rm q-m} &=& \frac{\Lambda A}{k_B T^2} \int_{\omega_0}^{4\omega_F+\omega_0} d\omega K(\omega) \omega^2 \sinh^{-2}\left(\frac{\omega}{2 k_B T}\right). ~~~~
 \end{eqnarray}
The remaining integral cannot be evaluated analytically, but we can study its different limiting cases. We get
\begin{widetext}
\begin{subequations}
\begin{eqnarray}
G_{\rm q-m} &=& \Lambda A k_B 
\frac{\sqrt{\pi} e^{-\omega_0/(k_B T)}\left[8 \omega_0^3 + 12 k_B T \omega_0^2 +18 (k_B T)^2 \omega_0 + 15 (k_B T)^3\right]}{4 (k_B T)^{3/2} \omega_F^{1/2}}, \quad \text{for } k_B T \ll \omega_0 \ll \omega_F   \label{eq:GqmN}\\
&=& \Lambda A k_B L_0 \frac{(k_B T)^{3/2}}{\omega_F^{1/2}},\quad \text{for }  \omega_0 \ll k_B T \ll  \omega_F \label{eq:GqmNa1}\\
&=& \Lambda A k_B \frac{\pi [\omega_0 + 2 \omega_F) (12 (k_B T)^2 - 10 \omega_F^2 - 4 \omega_0 \omega_F -\omega_0^2]}{3 (k_B T)^2}, \quad \text{for } \omega_0, \omega_F \ll k_B T \label{eq:GqmNa2}.
\end{eqnarray}
\end{subequations}
\end{widetext}
Here $L_0 = \int_0^\infty dx x^{5/2}[\cosh(x)-1]^{-1} \approx 8.91647.$
Note that as a function of temperature, $G_{\rm q-m}$ is monotonically increasing, but it saturates when $k_B T \gg \omega_F$. On the other hand, with respect to both the magnon band-gap $\omega_0$ and the Bloch-Gr\"uneisen type parameter $\omega_F$ the behavior is non-monotonous when $\omega_0 \ll \omega_F$, with a maximal value obtained when $\omega_{F/0}$ is of the order of $k_B T$. 

To compare the electron-magnon heat conductance $G_{\rm q-m}$ of the thin film N-FI
with the electron-phonon heat conductance, we here consider the bulk  electron-phonon 
heat conductance of N as \cite{wellstood},
\begin{eqnarray}
G_{\rm q-ph} = 5\Sigma \Omega T^4, \label{qmag6}
\end{eqnarray} 
where $\Sigma$ is the material dependent electron-phonon coupling constant and $\Omega$ is the volume of the quasi two-dimensional N. 
Comparing the analytical estimate of $G_{\rm q-m}$ in Eqs.~\eqref{eq:GqmN}-\eqref{eq:GqmNa2} with $G_{\rm q-ph}$ in Eq.~\eqref{qmag6}, we conclude that for small magnon band gaps at relatively low temperatures where $\omega_0\ll k_BT \ll \omega_F$,
the electron-magnon heat conductance can dominate over the
electron-phonon mechanism, whereas at high temperatures 
the electron-phonon heat conductance is the dominant thermalization mechanism.
The relative importance of these two processes changes at a crossover temperature, where both  heat conductances are equal to each other. 
Note that $G_{\rm q-ph}$ in Eq.~\eqref{qmag6} is obtained after assuming a continuous spectrum of 
three-dimensional wave vectors,
whereas for the electron-magnon heat conductance we include only a two-dimensional integral. The latter is primarily due to the fact that in the 
N-FI bilayer the electron-magnon coupling is a surface effect, and secondarily due to our assumption of thin films. In thin films then Eq.~\eqref{qmag6} overestimates the actual electron-phonon heat conductance and underestimates the crossover temperature.
In addition, the interface could in principle have some dynamical modes (say, some charges hopping from one place to another),
but these would have to connect to the continuum to realize a full heat conductance for bulk materials. They hence do not form a new channel, but can modify the coupling constants. We here disregard such effects due to their non-generic nature.

Motivated by the detector application, we also study the electron-magnon heat transport for the quasi two-dimensional S-FI hybrid structure. The kernel term in this case is (see the  discussion in Appendix \ref{DCT})
\begin{widetext}
\begin{eqnarray}
  K^{(\pm)}(E,\omega) &=&  \sqrt{\frac{\omega}{\omega_F}}\sqrt{\frac{\omega}{\omega -\omega_0}}  
    N_S(E\mp h) N_S(E\pm h\mp\omega) 
  \Theta (E)  \nonumber \\
  && \times \left[ 1 +   \Theta(E\pm h\mp\omega)\frac{\Delta^2}{(E\mp h)(E\pm h\mp\omega)}   \right] 
    \left[4-\left(\frac{\omega - \omega_0}{\omega_F}\right)\right]^{-1/2},
  \label{em6c}
\end{eqnarray}
where $N_S(E)=\left|\mathrm{Re}\left[{(E+i\Gamma)}/{\sqrt{(E+i\Gamma)^2-\Delta^2}} \right]\right|$
is the reduced superconducting density of states, $\Gamma\ll\Delta$
is the Dynes parameter \cite{dynes84} and $\Theta(x)$ is the Heaviside function. Note that Eq.~\eqref{em6c} couples  the two different spin components of the superconducting density of states. This is due to the spin-flip mechanism via electron-magnon interaction, as Eq.~\eqref{em4} represents.
Now, using Eqs.~\eqref{emag1}-\eqref{emag4} and \eqref{em6c} and
considering $\omega_0=h=0$, $k_BT < \Delta \ll 2\omega_F$ we analytically estimate the electron-magnon heat conductance of S-FI film as (see the derivation in Appendix \ref{emana})
\begin{eqnarray}
G_{\rm q-m} &=& \frac{k_B^{5/2}T^{3/2}\Lambda A}{\sqrt{\omega_F}} \left(\tilde{\Delta}e^{-\tilde{\Delta}}\sum_{n=0}^{\infty}\frac{D_n}{\tilde{\Delta}^n}   
+ \tilde{\Delta}^{5/2}e^{-2\tilde{\Delta}} \sum_{n=0}^{\infty}\frac{E_n}{\tilde{\Delta}^n}  \right), \label{march171}
\end{eqnarray}
where $\tilde{\Delta}=\Delta
/k_BT$. The lowest-order coefficients are $D_0=4.82$,
$D_1=2.88$, $E_0=\sqrt{2}\pi$ and $E_1=\pi/\sqrt{2}$.
The two sums in Eq.~\eqref{march171} are for  
quasiparticle-magnon scattering and magnon driven quasiparticle recombination, respectively. Contrary to
the electron-phonon heat conductance discussed below, the
scattering term dominates at all temperatures, so the recombination
term can also be disregarded to the first approximation.
The analytical estimate reveals the dominant exponential decay of
$G_{\rm q-m}$ at low temperatures $k_BT\ll \Delta$.
As $k_BT$ approaches $\Delta$, $G_{\rm q-m}$ 
follows a linear combination of different power laws as a function of temperature.
\end{widetext}

We compare the electron-magnon heat conductance of the thin film S-FI 
with the electron-phonon heat conductance of the superconductor, obtained from
\cite{Tero_bolometer, Tero_2017,heikkila2019thermal}
\begin{subequations}
\begin{eqnarray}
&& G_{\rm q-ph} = \frac{\Sigma\Omega}{96\zeta(5)k_B^6T^2}\int_{-\infty}^{\infty} dE~E\int_{-\infty}^{\infty}d\omega~
\omega^2|\omega|  \nonumber \\
&&~~~~~~~~~~~ \times L_{E,E+\omega}F_{E,\omega}, \label{res4} 
\end{eqnarray}
\begin{eqnarray}
&&\hspace{-7mm} L_{E,E^\prime} = \frac{1}{2}\sum_{\sigma=\uparrow,\downarrow}N_{\sigma}(E)N_{\sigma}(E^\prime) \nonumber \\
&&\hspace{-4mm}~~~~~~~~ \times \left[ 1-\Delta^2/[(E+\sigma h)(E^\prime +\sigma h)] \right],  \label{res5}  
\end{eqnarray}
\begin{eqnarray}
&&\hspace{-7mm} F_{E,\omega} = -\frac{1}{2}\left[ \sinh\left(\frac{\omega}{2k_BT}\right)
\cosh\left(\frac{E}{2k_BT}\right) \right. \nonumber \\
&&\hspace{-4mm}~~~~~~~~~~ \times \left. \cosh\left(\frac{E+\omega}{2k_BT}\right)  \right]^{-1}, \label{res6}
\end{eqnarray}
\end{subequations}
where $\Sigma$ is the material dependent electron-phonon coupling constant, $\Omega$ is the volume
of the film, $N_\sigma (E)=N_S(E+\sigma h)$ where $\sigma=\pm 1$ for $\sigma=\uparrow/\downarrow$,
and $\zeta(5)$ is the Riemann zeta function. The analytical estimate of the bulk value of $G_{\rm q-ph}$ is \cite{Tero_bolometer, Tero_2017}
\begin{eqnarray}
&&G_{\rm q-ph} \approx \frac{\Sigma\Omega}{96\zeta(5)}T^4\left[ \cosh(\tilde{h})e^{-\tilde{\Delta}} f_1(\tilde{\Delta}) \right. \nonumber \\
&&~~~~~~~~~~~ \left. + \pi\tilde{\Delta}^5e^{-2\tilde{\Delta}} f_2(\tilde{\Delta})\right],  \label{res10}
\end{eqnarray}
where $\tilde{h}=h/k_BT$ and $\tilde{\Delta}=\Delta/k_BT$. In Eq.~\eqref{res10} the terms $f_1$ and $f_2$
represent the scattering and recombination processes. The latter
dominates over the previous for $k_B T \gtrsim 0.1 \Delta$ and vice
versa, so both terms need to be taken into account. The functions 
$f_1(x)=\sum_{n=0}^3 C_n/x^n$ and $f_2(x)=\sum_{n=0}^2 B_n/x^n$, where
$C_0=440$, $C_1=-500$, $C_2=1400$, $C_3=-4700$, $B_0=64$, $B_1=144$, $B_2=258$.
Comparing the two analytical estimates, for $G_{\rm q-m}$ in
Eq.~\eqref{march171} and for $G_{\rm q-ph}$ in Eq.~\eqref{res10}, we note
that the electron-magnon
thermalization process can dominate the electron-phonon process at low
temperatures, whereas electron-phonon is the dominating mechanism at
high temperatures. As a result there can be a crossover temperature, where both heat conductances are equal to each other.
Here also it is important to note, as in the case without superconductivity, that Eqs.~\eqref{res4}-\eqref{res6} and \eqref{res10} are obtained assuming the continuous spectrum of three-dimensional wave vectors 
of the superconducting electrons,
whereas for the electron-magnon heat conductance we include only a
two-dimensional integral. As a result this overestimates $G_{\rm
  q-ph}$ and underestimates the resulting crossover temperature.

\section{Results and discussions}

In what follows we numerically analyze the electron-magnon heat conductance of the thin film normal metal-ferromagnetic insulator, N-FI,
and the thin film superconductor-ferromagnetic insulator, S-FI,
hybrid structures. We also compare the electron-magnon heat conductance with the bulk electron-phonon
heat conductance in the absence and in the presence of superconductivity. 

\subsection{Normal metal-ferromagnetic insulator}
\label{sec:N-FI}
\begin{figure}
\includegraphics[width=8cm]{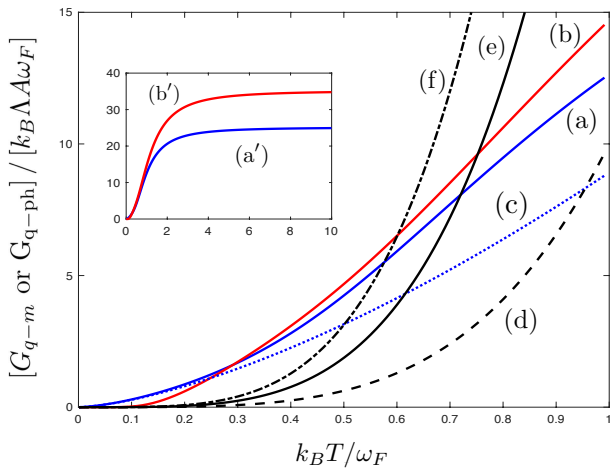}
\caption{\label{NFI} 
Electron-magnon $G_{\rm q-m}$ [Eqs.~(\eqref{emag3}-\eqref{emag5}), curves (a)-(c) and (a'), (b')] and electron-phonon heat conductance
$G_{\rm q-ph}$ [Eq.~\eqref{qmag6}, curves (d)-(f)] vs temperature $T$ for the
thin film normal metal-ferromagnetic insulator hybrid structure. In
(a) and (b) the chosen magnon band gap is $\omega_0=0$ and
$\omega_0=0.8\omega_F$, respectively. Curve (c) is the analytical estimate of electron-magnon heat conductance from Eq.~\eqref{eq:GqmNa1}, 
valid at $k_B T \ll \omega_F$ for the magnon band gap $\omega_0=0$. The curves $\mathrm{(a^\prime)}$ and $\mathrm{(b^\prime)}$ in the 
inset represent the corresponding extended plots of (a) and (b),
respectively. Curves (d)-(f) show the electron-phonon heat conductance
for three different thicknesses of the normal-metal film, for $d_N/d_{l}=2$, $d_N/d_{l}=6$ and $d_N/d_{l}=10$, respectively.}
\end{figure}

Here we first discuss the electron-magnon heat conduction in a thin film N-FI hybrid structure.
In Fig.~\ref{NFI}, we plot the electron-magnon heat conductance $G_{\rm q-m}$ vs
temperature $T$, for various magnon band gaps $\omega_0$ and compare with the analytical estimate
of $G_{\rm q-m}$ for $\omega_0=0$.
In line with Eqs.~\eqref{eq:GqmN}-\eqref{eq:GqmNa2}, we find that $G_{\rm q-m}$ 
decreases exponentially with a decreasing $T$ for $k_BT\ll \omega_0$,
and reaches a constant value, $4\pi k_B\Lambda A (\omega_0 + 2\omega_F)$,  for $k_BT\gg \omega_F$.
Figure \ref{NFI} also contains the bulk electron-phonon heat conductance
$G_{\rm q-ph}$ of the thin film N vs $T$
for various film thicknesses $d_N$ of the normal metal. 
To find out the relative importance between electron-magnon and the
usual electron-phonon thermalization mechanisms, we now compare $G_{\rm q-m}$ with $G_{\rm q-ph}$.
For the comparison, we define a crossover temperature $T^*$, 
where the $G_{\rm q-ph}$ vs $T$ curve crosses the $G_{\rm q-m}$ vs $T$ curve.
At the characteristic temperature we thus have 
\begin{equation}
G_{\rm q-m}(T=T^*)=G_{\rm q-ph}(T=T^*). \label{Tstar1a}
\end{equation}
Since the electron-magnon heat conduction is an interface process, and
the electron-phonon conduction is a bulk process, the crossover
temperature depends on the normal metal thickness $d_N$. As expected,
we can see from Fig.~\ref{NFI} that the electron-magnon process
dominates below $T^*$ and vice versa for the electron-phonon process. 
Using Eqs.~\eqref{eq:GqmNa1} and \eqref{qmag6}, we get 
\begin{figure}
\includegraphics[width=8cm]{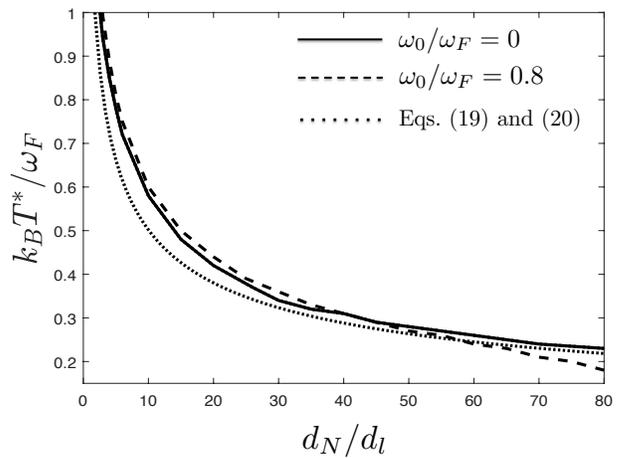}
\caption{\label{NFI_Tstar} 
Temperature $T^*$ below which the electron-magnon thermalization dominates
the electron-phonon mechanism, as a function of the thickness $d_N$ of
the normal metal film. The curves are for two different values of the
magnon band gap. The dotted line shows the analytical
estimate from Eq.~\eqref{Tstar2a}. The thickness is scaled by $d_{l}$
defined in Eq.~\eqref{march74}.}
\end{figure}
\begin{eqnarray}
k_B T^* &=& 1.26 ~\omega_F \left(d_{l}/d_{N}\right)^{2/5}, \label{Tstar2a} \\
\mathrm{with}~d_{l} &=& k_B^5(\Lambda/\Sigma)\omega_F^{-3} \label{march74}
\end{eqnarray}
for $\omega_0=0$ and $k_BT^* < \omega_F$. This estimate works quite well even for a non-zero
$\omega_0$, as shown in Fig.~\ref{NFI_Tstar}. 
       
Let us also estimate typical values of parameters and the resulting
$T^*$. In particular, we consider EuS/Al and EuO/Al hybrid structures, where EuS and EuO are ferromagnetic insulators, and Al is a metal.
The Al characteristic electron-phonon coupling constant $\Sigma=0.2 \times 10^9$ Wm$^{-3}$K$^{-5}$
\cite{Tero_2006}. Both EuS and EuO are characterized by the effective lattice spin $s=7/2$ \cite{Rohling_2018},
lattice constant $\zeta=5.1$ $\mathrm{\AA}$ \cite{Rohling_2018} and
the characteristic (Bloch-Gr\"uneisen type) magnon frequency
$\omega_F/k_B=T_F=53$ K \cite{Rohling_2018}.
The interfacial coupling energy is around  $J=10$ meV \cite{Rohling_2018}, for
both hybrid structures, EuS/Al and EuO/Al, but its precise value depends on the quality of the contact.
Using these values and $\mu=\hbar^2 k_F^2/(2 m_e)$ with a free electron mass $m_e$, we get $\Lambda=7.5 \times 10^{48}$ J$^{-1}$m$^{-2}$s$^{-1}$ and 
$d_{l}=50$ pm. As a result we get the crossover temperature $T^*=3$ K for both hybrid structures with the Al thickness of $100$ nm.
Electron-magnon thermalization hence becomes relevant in modern-day low-temperature
experiments on thin film bilayers.

\subsection{Superconductor-ferromagnetic insulator structure}

Because many functionalities of low-temperature devices \cite{Tero_2006,Tero_2017,heikkila2019thermal} employ
superconductivity, we also analyze the effect of superconductivity on
the electron-magnon heat conduction. In this case two new energy
scales show up:  the superconducting energy gap $\Delta$ and the
exchange field $h$ induced by the magnetic proximity effect into the
superconductor \cite{Meservey,heikkila2019thermal}. The latter might
be present also in the normal state, but there it is not relevant to
the magnitude of the heat conductance as long as it is much smaller
than $\mu$. 

\begin{figure}[h]
\includegraphics[width=8cm]{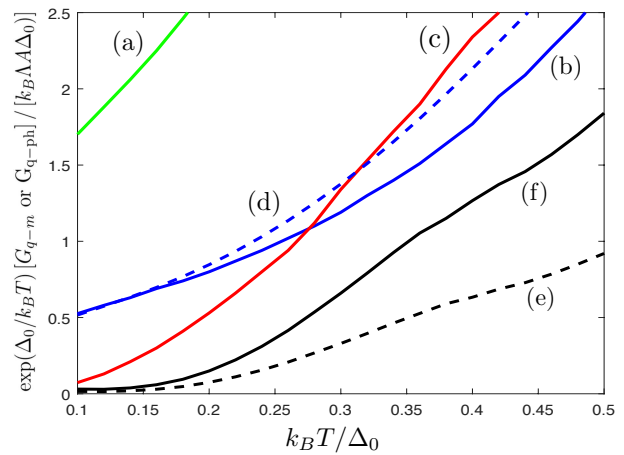}
\caption{\label{SFI_final3} 
Electron-magnon $G_{\rm q-m}$ [Eqs.~(\eqref{emag3}, \eqref{emag4} and \eqref{em6c}), curves (a)-(d)] and 
electron-phonon heat conductance $G_{\rm q-ph}$ [Eqs.~(\eqref{res4}-\eqref{res6}), curves (e)-(f)] vs temperature $T$ for the
thin film superconductor-ferromagnetic insulator hybrid
structure.
The parameters for the curves are (a) $\omega_0=0$, $\omega_F=\Delta_0$ and
(b) $\omega_0=0$, $\omega_F=10\Delta_0$ and (c) $\omega_0=0.5\Delta_0$, $\omega_F=10\Delta_0$.
The curve (d) is the analytical estimate (Eq.~\eqref{march171}) of the electron-magnon heat conductance
for the given parameters in (b) curve. 
The thin film superconductor thicknesses are in curve (e) $d_S/d_\Delta = 0.4$
and (f) $d_S/d_\Delta = 0.8$.
For all the curves we set $h=0$ and $\Gamma=10^{-3}\Delta_0$.}
\end{figure}

\begin{figure}[h]
\includegraphics[width=8cm]{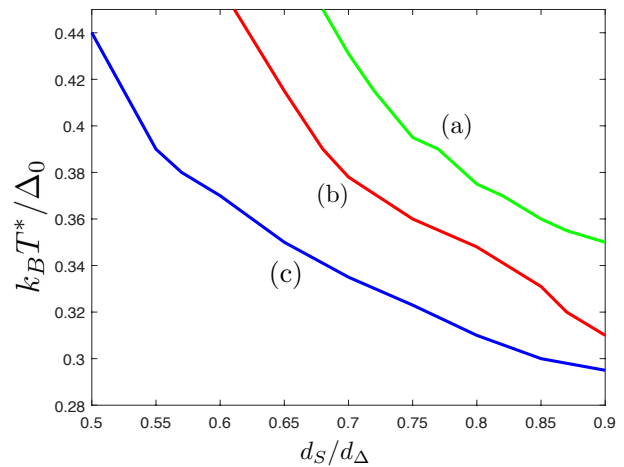}
\caption{\label{Tstar_SFI_Dec05_2018} Crossover temperature $T^*$
  below which electron-magnon thermalization becomes dominant, as a
  function of the thickness $d_S$ of the superconductor. For all curves we set $h=0$ and $\Gamma=10^{-3}\Delta_0$.
The parameters for the curves are (a) $\omega_0=0$, $\omega_F = 26\Delta_0$ (b) $\omega_0=0.5\Delta_0$, $\omega_F = 50\Delta_0$ and 
(c) $\omega_0=0$, $\omega_F = 50\Delta_0$.}
\end{figure}

Since the superconducting gap $\Delta (T,h)$ depends on $h$ and $T$, 
in what follows we introduce scaling energy as the magnitude of the gap $\Delta_0$ at $T=0$ K and $h=0$.
We compute $\Delta (T,h)$ self-consistently using Eq.~\eqref{selfdelta1} (see Appendix \ref{Selfconsistent}).
Self-consistent calculation is significant near the critical magnetic field \cite{Chandrasekhar_1962,Clogston_1962},
and near the critical temperature, but does not otherwise affect the results much. 
In Fig.~\ref{SFI_final3} we plot again the two heat conductances $G_{\rm q-m}$ and $G_{\rm q-ph}$ in
the case where the metal is in the superconducting state.
As in the analytical estimates, Eqs.~\eqref{march171} and \eqref{res10}, both decay
exponentially at low temperatures $k_B T\ll \Delta$ due to the
exponential decay of the number of quasiparticles, $\sim
\exp(-\Delta/k_B T)$. It is thus easier to compare their ratio, or the
temperature $T^*$ at which they become equal. That temperature is plotted in
Fig.~\ref{Tstar_SFI_Dec05_2018}. 
We can see that the overall behavior
with respect to $d_S$ is quite similar to the normal state, but 
superconductivity affects the two processes slightly differently.
In  Figs.~\ref{SFI_final3} and \ref{Tstar_SFI_Dec05_2018}, we  introduce a length scale
\begin{eqnarray}
d_{\Delta} = k_B^5 (\Lambda/\Sigma)\Delta_0^{-3}, \label{march26}
\end{eqnarray}
associated with 
scaling energy $\Delta_0$. Note that in the usual case $\omega_F \gg \Delta$, $d_\Delta \gg d_l$ introduced in Eq.~\eqref{march74}. In order to get the crossover temperature $T^*$ to be significantly below the superconducting critical temperature $T_c$, we would hence have to assume thicker films or smaller exchange couplings than those discussed in the previous section. Besides $\omega_F$ and $\Delta$, also the precise value of the magnon band gap affects $T^*$. However, we find that $T^*$
is slowly varying with the superconductor film thickness ($d_S$) irrespective 
of the small magnon band gap in Fig.~\ref{Tstar_SFI_Dec05_2018}. 

In Sec.~\ref{sec:N-FI}, we find that an EuS/Al film with 100 nm Al layer can have a crossover temperature at 3 K, 
much above the Al $T_c$ (usually $1. 2$ K in thin films in absence of spin-splitting field). 
Hence, to find the crossover in EuS/Al films in the superconducting state, the Al layer 
should be much thicker. Using the parameters of EuS/Al and EuO/Al as i
n Sec.~\ref{sec:N-FI} with $\Delta_0/k_B=2$ K \cite{Matthias}, we get $d_\Delta=900$ nm. 
Hence $d_S=100$ nm would correspond to $ T^* > T_c$, consistent with the normal-state estimate. 

\begin{figure}[h]
\includegraphics[width=8cm]{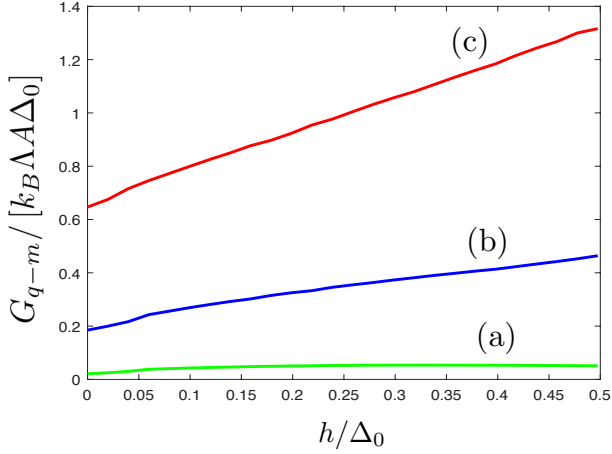}
\caption{\label{SFI_h} 
    Electron-magnon heat conductance $G_{\rm q-m}$ vs spin-splitting field $h$ for the
thin film superconductor-ferromagnetic insulator hybrid structure.
For all curves 
$\omega_0=0.1\Delta_0$, $\omega_F=\Delta_0$ and $\Gamma=10^{-3}\Delta_0$.
The curves (a), (b) and (c) are for $k_BT=0.2\Delta_0$, $k_BT=0.3\Delta_0$
and $k_BT=0.4\Delta_0$, respectively.}
\end{figure}

Let us next study the effect of the induced spin-splitting field on the
electron-magnon heat conductance. That is plotted in
Fig.~\ref{SFI_h} at three different temperatures for a low value
of the magnon gap $\omega_0$. Note that we here neglected the effect of spin
relaxation, which may become especially relevant for higher
$h$. Perhaps surprisingly, the effect of the spin splitting on $G_{\rm q-m}$ is quite modest, taking into account that the field reduces the
energy gap from $\Delta$ to $\Delta-h$ for one of the spin
species. However, since the electron-magnon coupling couples both
spins, this reduced gap is not immediately visible. 

\begin{figure}[h]
\includegraphics[width=8cm]{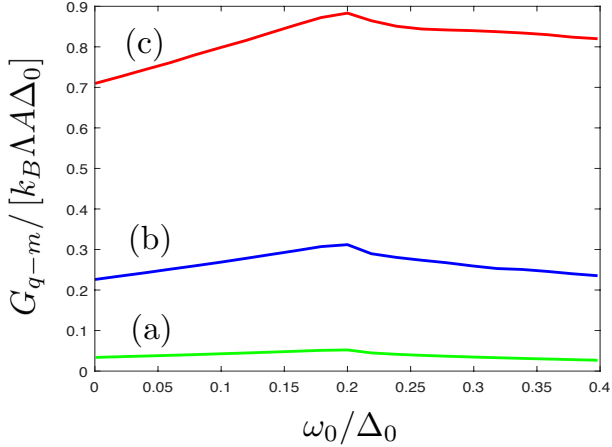}
\caption{\label{SFI_w0}
    Electron-magnon heat conductance $G_{\rm q-m}$ vs magnon band gap $\omega_0$ for the
thin film superconductor-ferromagnetic insulator hybrid structure.
For all the curves 
$h=0.1\Delta_0$, $\omega_F=\Delta_0$ and $\Gamma=10^{-3}\Delta_0$.
The curves (a), (b) and (c) are for $k_BT=0.2\Delta_0$, $k_BT=0.3\Delta_0$
and $k_BT=0.4\Delta_0$, respectively. }
\end{figure}

In the superconducting case, also the relation between the
spin-splitting field and the magnon gap $\omega_0$ affects the
magnitude of the electron-magnon heat
conductance. This is shown in Fig.~\ref{SFI_w0} showing $G_{\rm q-m}$ as a
function $\omega_0$ for $h=0.1\Delta_0$. When $\omega_0 \approx 2h$,
$G_{\rm q-m}$ has a shallow maximum (a kink), as this is where the magnons just
above the gap edge couple the electrons at the edges of the two spin
bands (see Eq.~\eqref{em6c}). However, due to the low density of states of the magnons at the
gap edge, the dependence is not very strong. It might however be
observable in the case where the spin-splitting field is tuned with an
external magnetic field. Hence especially in the superconducting case
the electron-magnon heat conductance can be used to obtain
spectroscopic information about the magnons. 

\section{Conclusions}

In this work we have studied heat transport between the electrons in a metallic thin film in a normal or a superconducting state and the magnons in a nearby
 ferromagnetic insulator film, resulting from the interfacial electron-magnon interaction. This mechanism can dominate
over the electron-phonon heat transport at low temperatures and hence
should be taken into account in device concepts \cite{Tero_2017} utilizing such hybrid
structures at low temperatures. The crossover temperature below which the
electron-magnon process starts to dominate depends on the properties
of the magnet and naturally on the electron-magnon interaction, but
also on the thickness of the metal film. For reasonable values of the parameters of these films 
we find that this crossover temperature can be of the order of $1$ Kelvin.
In this work we assume
that the magnons flow away and only somewhere far from the interface thermalize with the phonons.
In this situation the extra heat resistance related to this thermalization mechanism can be disregarded. Similarly, depending on the device geometry, one might have to include the (Kapitza) thermal boundary resistance for thermalizing the phonons, and this would affect the overall heat balance and the crossover temperatures. 
In the superconducting state, the magnitude of the induced spin-splitting
field also affects the size of the heat conductance. In particular,
the heat conductance obtains a maximum when the spin-splitting field
equals half of the gap in the magnon spectrum. Because the
spin-splitting field can be varied by using an external field (see,
e.g., \cite{xiong2011}), this dependence can be studied in
detail. Such a study would hence reveal spectroscopic information
about the magnon spectrum in the ferromagnetic insulator.  

This project was supported by the Academy of Finland via its Key Funding project
 (Project No. 305256) and regular project number 317118 and from the European
 Union's Horizon 2020 research and innovation programme under grant
 agreement No 800923 (SUPERTED).

\appendix

\section{Discrete to continuous transformation}
\label{DCT}

Here we demonstrate the discrete to continuous transformation for the case of the N-FI hybrid structure. 
For thin films $\vec{k}$ and ${\vec{q}}$ are two-dimensional,
and hence we have the following discrete to continuous transformation
\begin{widetext}
\begin{eqnarray}
 \sum_{\vec{k},\vec{q}} F(\epsilon_{\vec{k}},\omega_{\vec{q}}) \delta (\epsilon_{\vec{k}+\vec{q}}-\epsilon_{\vec{k}}\pm \omega_{\vec{q}}\mp 2h) 
&=&  \left(\frac{A}{4\pi^2}\right)^2 \int d^2k~d^2q ~F(\epsilon_{\vec{k}},\omega_{\vec{q}}) 
 \delta (\epsilon_{\vec{k}+\vec{q}}-\epsilon_{\vec{k}}\pm \omega_{\vec{q}}\mp 2h)  \nonumber \\
&= &\left( \frac{\pi k_F^2A^2}{32\pi^4B\mu} \right) \int_{0}^{\infty} d\epsilon_{\vec{k}}\int_{\omega_0}^\infty d\omega_{\vec{q}} \int_0^{2\pi} d\theta
~ F(\epsilon_{\vec{k}},\omega_{\vec{q}}) 
 \delta (\epsilon_{\vec{k}+\vec{q}}-\epsilon_{\vec{k}}\pm \omega_{\vec{q}}\mp 2h). ~~~
\label{A1}
\end{eqnarray} 
To obtain Eq.~\eqref{A1} we have used the energy dispersion of the normal-metal electrons, $\epsilon_{\vec{k}}=\mu k^2/k_F^2$, 
and the energy dispersion of the magnons, $\omega_{\vec{q}}=\omega_0 + Bq^2$.
Therefore we have $kdk=k_F^2/(2\mu)d\epsilon_{\vec{k}}$ and $qdq=1/(2B)d\omega_{\vec{q}}$.
We also have
\begin{eqnarray}
\epsilon_{\vec{k}+\vec{q}}-\epsilon_{\vec{k}}\approx \hbar v_F\sqrt{\frac{\omega_{\vec{q}} -\omega_0}{B}}\cos\theta +\frac{ \mu(\omega_{\vec{q}} -\omega_0)}{Bk_F^2},
\label{A2}
\end{eqnarray}
where $\theta$ is the angle between $\vec{k}$ and $\vec{q}$,
and the Fermi energy $\mu$ is much larger than the relevant magnon energies $\omega_{\vec{q}}$.
Here after integrating the Dirac delta function over $\theta$ in Eq.~\eqref{A1}, we obtain the following result,
\begin{eqnarray}
&& \int_0^{2\pi} d\theta~ \delta (\epsilon_{\vec{k}+\vec{q}}-\epsilon_{\vec{k}}\pm \omega_{\vec{q}}\mp 2h) 
=   \frac{1}{\mu} \sqrt{\frac{\omega_F}{\omega_{\vec{q}}-\omega_0}} 
 \left|\mathrm{Re}\left[1-\frac{1}{4}\left(\frac{\omega_{\vec{q}} - \omega_0}{\omega_F}\right)
\right]^{-1/2}\right|,
 \label{A3}
\end{eqnarray}
\end{widetext}
assuming the relevant magnons and the weak spin-splitting field 
satisfy $(\omega_{\vec{q}}\pm 2h)/\mu \rightarrow 0$.
Equations \eqref{A1}-\eqref{A3} are used in Eqs.~\eqref{emag1}-\eqref{emag5}.
 
Next, we have followed the similar mathematical protocol in the case
where the metal becomes superconducting. 
In this case $\left|{dE_{\vec{k}+\vec{q}}}/{d\theta}\right|=2\mu |\sin\theta|/N_S(E_{\vec{k}+\vec{q}})$
and $d\epsilon_{\vec{k}}=N_S\left(E_{\vec{k}}\right)dE_{\vec{k}}$, where $E_{\vec{k}}=\sqrt{\left(\epsilon_{\vec{k}}-\mu\right)^2+\Delta^2}$.
Here $N_S$ is the superconducting density of states.
After the discrete to continuous transformation and integrating the
Dirac delta functions analogous to that above, we get the 
kernel terms $K^{(\pm)}$ in Eq.~\eqref{em6c}. 

\section{Electron-magnon heat conductance of S-FI at low temperatures}
\label{emana}

In order to obtain the analytical expression of $G_{\rm q-m}$ of a S-FI hybrid structure we here
consider $\omega_0=h=0$ and $k_BT<\Delta \ll 2\omega_F$, such that we can effectively have $4\omega_F/k_BT \rightarrow \infty$ and $\omega/4\omega_F \rightarrow 0$.
Now using Eqs.~\eqref{emag4} and \eqref{em6c} we have
\begin{eqnarray}
&& G^{(\pm)}_{\rm q-m} = \pm \frac{k_B^{5/2}T^{3/2}}{8\sqrt{\omega_F}} \int_0^\infty dx \int_0^\infty dy~ y^{3/2} \tilde{N}_S(x)\tilde{N}_S(x\mp y) \nonumber \\
&&~~~~~~~~~~~ \times \left[ 1 + \Theta(x\mp y)\frac{\tilde{\Delta}^2}{xy} \right] F(x,y),  \label{B1} 
\end{eqnarray}
\begin{eqnarray}
&& F(x,y) =
\mathrm{cosech}\left(\frac{y}{2}\right)
\mathrm{sech}\left(\frac{x}{2}\right)\mathrm{sech}\left(\frac{x\mp y}{2}\right), \label{B2}
\end{eqnarray}
where $\tilde{N}_S(x)=\lim_{\tilde{\Gamma}\rightarrow 0}\left|\mathrm{Re}\left[{(x+i\tilde{\Gamma})}/{\sqrt{(x+i\tilde{\Gamma})^2-\tilde{\Delta}^2}} \right]\right|$ and $\tilde{\Delta}=\Delta/k_BT$.
The integrand in Eq.~\eqref{B1} is nonzero only for $x\ge \tilde{\Delta}$
and $x\mp y \ge \tilde{\Delta}$, hence at low temperatures we can approximate
\begin{eqnarray} 
&& \mathrm{sech}\left(\frac{x}{2}\right)\approx 2e^{-|x|/2}, \label{B3} \\
&& \mathrm{sech}\left(\frac{x\mp y}{2}\right)\approx 2e^{-|x\mp y|/2}. \label{B4}
\end{eqnarray}
Combining Eqs.~\eqref{B1}-\eqref{B4} we obtain
\begin{widetext}
\begin{eqnarray}
&& G^{(+)}_{\rm q-m}-G^{(-)}_{\rm q-m} = \frac{k_B^{5/2}T^{3/2}}{8\sqrt{\omega}_F} \int_{\tilde{\Delta}}^\infty dx \int_{\tilde{\Delta}}^\infty dy~
\frac{xy + \tilde{\Delta}^2}{\sqrt{(x^2-\tilde{\Delta}^2)(y^2-\tilde{\Delta}^2)}} F_1(x,y) \nonumber \\
&&~~~~~~~~~~~~~~~~~~~~~  + \frac{k_B^{5/2}T^{3/2}}{8\sqrt{{\omega}_F}} \int_{\tilde{\Delta}}^\infty dx  \int_{-\infty}^{-\tilde{\Delta}}dy~
\tilde{N}_S(x)\tilde{N}_S(y) F_1(x,y), \label{B5}   \\
&& F_1(x,y) = 4 |x-y|^{3/2} \mathrm{cosech} (|x-y|/2) e^{-|x|/2}e^{-|y|/2}.   \label{B6}
\end{eqnarray}
The first term in the right hand side in Eq.~\eqref{B5} represents
quasiparticle-magnon scattering, where as the second term is due to
quasiparticle recombination processes.
Now approximating $\sinh^{-1}\left(\frac{2\tilde{\Delta}+x+y}{2}\right)=2e^{-\tilde{\Delta}}e^{-(x+y)/2}$ for $x,y >0$,
we finally have
\begin{eqnarray}
&& G^{(+)}_{\rm q-m} - G^{-}_{\rm q-m} = \frac{k_B^{5/2}T^{3/2}}{\sqrt{{\omega}_F}} \left(\tilde{\Delta}e^{-\tilde{\Delta}}\sum_{n=0}^{\infty}\frac{D_n}{\tilde{\Delta}^n}   
+ \tilde{\Delta}^{5/2}e^{-2\tilde{\Delta}} \sum_{n=0}^{\infty}\frac{E_n}{\tilde{\Delta}^n}  \right), \label{B7} \\
\implies && G_{\rm q-m} =  \frac{k_B^{5/2}T^{3/2}\Lambda A}{\sqrt{{\omega}_F}} \left(\tilde{\Delta}e^{-\tilde{\Delta}}\sum_{n=0}^{\infty}\frac{D_n}{\tilde{\Delta}^n}   
+ \tilde{\Delta}^{5/2}e^{-2\tilde{\Delta}} \sum_{n=0}^{\infty}\frac{E_n}{\tilde{\Delta}^n}  \right), \label{B8}
\end{eqnarray}
with the lowest-order coefficients $D_0=4.82$,
$D_1=2.88$, $E_0=\sqrt{2}\pi$, $E_1=\pi/\sqrt{2}$.
\end{widetext}

\section{Self-consistent equation for $\Delta (T,h)$}
\label{Selfconsistent}

Neglecting spin relaxation effect, we have the self-consistent equation
for the superconducting gap, $\Delta (T,h)$, as
\begin{eqnarray}
&& \Delta =\frac{\lambda}{2}\int_{-\Omega_D}^{\Omega_D}d\epsilon\ \text{Im}\left[F_{01}(\epsilon)\right]\tanh\left(\frac{\epsilon}{2k_BT}\right),\label{selfdelta1}  
\end{eqnarray}
where $\lambda$ is the effective coupling constant, $\Omega_D$ is the Debye cutoff energy and 
\begin{eqnarray}
&& F_{01}(\epsilon) = \frac{1}{2} \left[ F_{0}(\epsilon +h) + F_{0}(\epsilon -h)  \right], \label{selfdelta2}  \\
&& F_0(\epsilon) = \frac{i \Delta }{\sqrt{(\epsilon+i\Gamma)^2-\Delta^2}}   \label{selfdelta3}
\end{eqnarray}
with the Dynes parameter $\Gamma$. We use this self-consistent superconducting gap to compute various quantities in the main text of the paper.

\bibliography{Reference} 

\end{document}